\documentclass[prd,twocolumn,showpacs,preprintnumbers]{revtex4}
\pdfoutput=1
\usepackage{amsfonts,amsmath,amssymb}
\usepackage{mhchem}
\usepackage{graphicx}
\usepackage{color}
\usepackage{natbib}
\usepackage{lipsum}
\usepackage{float}
\usepackage[plainpages=false, colorlinks=true, anchorcolor=blue, linkcolor=blue, citecolor=blue, bookmark
s=false]{hyperref}

\newcommand{\rthis}[1]{\textcolor{black}{#1}}

\begin{document}

\newcommand{\apjl}{Astrophys. J. Lett.}
\newcommand{\apjs}{Astrophys. J. Suppl. Ser.}
\newcommand{\aap}{Astron. \& Astrophys.}
\newcommand{\aj}{Astron. J.}
\newcommand{\araa}{Ann. Rev. Astron. Astrophys. } 
\newcommand{\mnras}{Mon. Not. R. Astron. Soc.}
\newcommand{\solphys}{Solar Phys.}
\newcommand{\jcap}{JCAP}
\newcommand{\pasj}{PASJ}
\newcommand{\pasa}{Pub. Astro. Soc. Aust.}
\newcommand{\apss}{Astrophysics \& Space Science}
\newcommand{\aaps}{Astron. Astrophys. Suppl. Ser.}
\newcommand{\Dtwoo}{$\mathrm{D_2O}$ }
\newcommand{\cl}{$\mathrm{^{36}Cl}$ }
\newcommand{\I}{$\mathrm{^{123}I}$ }
\newcommand{\tc}{$\mathrm{^{99m}Tc}$ }

\title{Generalized Lomb-Scargle analysis of   $\rm{^{123}I}$ and $\rm{^{99m}Tc}$ decay rate measurements}

\author{Gautham Gururajan}
\altaffiliation{ep17btech11008@iith.ac.in}
\author{Shantanu  Desai}
\altaffiliation{shntn05@gmail.com}
\affiliation{Department of Physics, IIT Hyderabad, Kandi, Telangana 502285 India}

\begin{abstract}

We apply the generalized  Lomb-Scargle periodogram to the  \I and \tc decay rate measurements based on  data taken at   the Bronson Methodist Hospital. The aim of this exercise  was to carry out an independent search for sinusoidal modulation for these radionuclei (to complement \rthis{the analysis in Borrello et al.}) at frequencies for which other radionuclei have shown periodicities.   We do not find  such a modulation at any frequencies, including annual modulation  or at frequencies associated with solar rotation. Our analysis codes and datasets have been made publicly available. 
\end{abstract}
                                                     
\maketitle

\section{Introduction}

Over the past two decades,  multiple groups (starting with Falkenberg~\cite{Falkenberg}),  have argued for periodicities in the beta decay rates for various radioactive nuclei. Periodicities have been \rthis{reported} at 1 year (associated with the Earth-Sun distance)~\cite{Jenkins2}; 28 days (associated with solar rotation)~\cite{Sturrock12,Sturrocksolar}, 29.5 days (associated with synodic lunar month)~\cite{Parkhomov}, etc.~\citet{Sturrock01} have also found sinusoidal modulations in the solar neutrino data at the same  frequencies. They have correlated these two sets of findings, and hence argued for  the influence of solar rotation on    the beta decay measurements. In addition to the  above claims for  a sinusoidal variation in the beta decay rates, correlations between  beta decay rates and other transient astrophysical observations have also been found such as solar flares~\cite{Jenkins}, and also the first binary neutron star merger  seen in gravitational waves, GW170817~\cite{FischbachGW}.  A review of some of these claims can be found in~\cite{Jenkins13,Sturrock13,Sturrock16,Sturrock18}.

However,  other groups have failed to confirm these results, while analyzing the same data, or offered more prosaic explanations for the variability observed in the decay rate measurements. A review of some of the rejoinders and counter-rejoinders can be found in~\cite{Jenkins10,Sturrock13,Sturrock16,Sturrock18,Kossert,Pommesolar,PommePLB,Pomme,Pomme19,pomme2018decay} and references therein. Other groups have also refuted the results related to  an association between  the beta decay rates and solar flares~\cite{bellotti2,Baudis}. However, the jury is still out on some of these claims~\rthis{eg. the correlation between the decay rates of $^{32}$Si and  $^{36}$Cl with GW170817~\cite{FischbachGW}, although no such correlations  were seen in the  decays of other nuclei, such as $^{44}$Ti, $^{60}$Co, and $^{137}$Cs~\cite{Breur}}. 
One impediment in reproducing some of these results, is that not all the beta-decay data and associated measurement errors have been made publicly available.
To independently verify some of these claims, we   have analyzed some of the beta-decay and solar neutrino data ourselves using robust statistical methods, for whatever data was accessible or made publicly available. Our analysis shows periodicities  associated with solar rotation and annual modulation, although with a lower significance than claimed in some of the original works~\cite{Desai16,Tejas,Dhaygude}.

All the radioactive nuclei claimed to exhibit sinusoidal modulations are beta-decay emitters. Until recently, there was no  study to check if any radionuclei which undergo  isomeric transitions show variability, \rthis{and only one study for nuclei undergoing electron capture~\cite{okeefe}}. To rectify this, Borrello et al~\cite{Borrello} (B18, hereafter) looked for periodicities in the decays of \I (half-life of about 13 h) and \tc (half-life of about 6~h). These radionuclides decay from electron capture and isomeric transition, respectively. Their decay chain is shown schematically in Fig.~1 and Fig.~2 of B18. These isotopes are widely used for clinical nuclear medicine purposes. Therefore, the widespread use of these isotopes in medical physics provides an another impetus to look for variability, since any deviation from a constant decay rate would also have implications for  clinical studies.  B18 applied the Lomb-Scargle periodogram to look for periodicities.
From their analysis, no statistical significant peaks indicative of sinusoidal variations were found. They also did not find any correlation between their observed decay rates and solar activity as well as K-indices, which characterize the instability of Earth's magnetic field.

In this work, we independently try to analyze the radioactive decay measurements in B18 (which were kindly provided to us by  J.~Borello) using the Generalized Lomb-Scargle periodogram~\cite{Lomb,Scargle,Kurster} to look for any periodicities. Since the previous history of this field has shown that multiple groups analyzing the same data have reached drastically different conclusions~\cite{Desai16,Tejas,Dhaygude}, it behooves us to reanalyze this data and calculate significance of any possible periodicity using  robust statistical techniques. We calculate the statistical significance of the most significant peak as well as other periods deemed interesting in literature, such as annual variation, solar rotation~\cite{Sturrockcomp,Sturrocksolar}, using multiple methods. \rthis{For this analysis, we use the same methodology as in our previous work~\cite{Dhaygude}.}

The outline of this paper is as follows. We briefly recap some details of the Lomb-Scargle periodogram and different methods of calculating the  false alarm probability in Sect.~\ref{sec:ls}. A brief summary of the results by B18 is discussed in Sect.~\ref{sec:prevanalysis}. 
Our own analysis   is described in Sect.~\ref{sec:analysis}. 
We conclude in Sect.~\ref{sec:conclusions}.

\section{Generalized Lomb-Scargle Periodogram}
\label{sec:ls}
The Lomb-Scargle (L-S)~\cite{Lomb,Scargle,Vanderplas,Vanderplas15,astroml}  periodogram is a well-known technique to look for  periodicities in unevenly sampled datasets. Its main goal is to determine the  frequency ($f$)  of a  periodic signal in a time-series dataset  $y(t)$
given by:
\begin{equation}
y(t)=a\cos(2\pi f t)+ b \sin(2 \pi f t).
\label{eq:yt}
\end{equation}
The L-S periodogram calculates the power as a function of frequency, from which one can assess the statistical significance at a given frequency.

For this analysis, we use the generalized (or floating-mean)  L-S periodogram~\cite{Kurster,Bretthorst}. The main difference with respect to the ordinary L-S periodogram is that an arbitrary offset is added to the mean values. More details on the differences are elaborated in ~\cite{Vanderplas,Vanderplas15} and references therein.
The generalized L-S periodogram has been shown to be more  sensitive than the normal one, for detecting peaks, when  the  sampling  of the data overestimates the mean~\cite{Vanderplas,Kurster,proc}.

To evaluate the  statistical significance of any peak in the L-S periodogram, we need to calculate its  false alarm probability (FAP) or $p$-value. A large number of metrics
have been developed  to estimate the FAP of peaks in the L-S periodogram~\cite{Scargle,Vanderplas,Pommemod,Pommesig}. We use most of these to calculate the FAP for our analysis.  We now briefly describe enumerate these \rthis{different} techniques. 
\begin{itemize}
    \item {\bf Baluev}

    This method  uses extreme value statistics for stochastic process, to compute an upper-bound of the FAP for the alias-free case. The analytical expression for the FAP using this method can be found in ~\cite{Baluev,Vanderplas}.
    
    \item {\bf Bootstrap}
    
    This method uses non-parametric bootstrap resampling~\cite{Vanderplas}. It computes  L-S periodograms on synthetic data for the same observation times. \rthis{The bootstrap is the most robust estimate of the FAP, as it makes very few assumptions about the periodogram distribution, and the observed times  also fully account for survey window effects~\cite{Vanderplas}.}
    \item {\bf Davies}
    This method is similar to the Baluev method, but  is not accurate at large false alarm probabilities, where it shows values greater than 1~\cite{Davies}.
    \item {\bf Naive}
    
        This method is  based on the {\it ansatz} that well-separated areas in the periodogram are independent.
    The total number of such independent frequencies depend on the sampling rate and total duration,  and more details can be found in  ~\cite{Vanderplas}. 
    \end{itemize}

Once the FAP is known, based on  any of the above methods, one can evaluate the  $Z$-score or  significance in terms of number of sigmas~\cite{Cowan11,Ganguly}, in case the FAP is very small. A rule of thumb for any peak to be interesting is that FAP is less than $0.05$. However for a peak to be statistically significant, its $Z-$score must be greater than 5$\sigma$.

\section{Recap of B18 and datasets used}
\label{sec:prevanalysis}
Here, we briefly summarize the analysis in  B18, wherein more details can be found.
Their experiments were performed at the Bronson Methodist Hospital in Michigan. \I (Iodine) was provided as sodium iodide crystals. The contamination from $^{125}$I  was deemed to be less than  12.4\%. \tc  (Technetium) was supplied as sodium pertechnetate in aqueous solution with about 0.9\% contamination from sodium chloride. More details about the apparatus and experimental procedure used for measuring the half-life can be found in B18. 
Half-life measurements were performed over   a two-year period from May 2012 to June 2014. The mean time interval between \I measurements was about 7 days 10 h, and the same for \tc was 3 days 20 hr. 

L-S analysis was then applied  to the measured half-life data for both the radionuclides. A search for statistically significant peaks was done for both the nuclei upto 600 days. For \I, the maximum significance occurs at a period of 23.5 days with $p$-value of 0.24. For \tc, the maximum significance occurs at a period of 8.77 days  with a $p$-value of 0.47. Therefore, no statistically significant peaks were seen. Then 95\% c.l. upper limits were set on a periodic variation of one year. B18 then examined the outliers in the data for correlation with environmental factors, power supply voltage as well as for any outbursts in solar activity. No such correlations were seen. Therefore, B18 concludes that the \I and \tc data show no periodic variations, with limits on the amplitude of  annual variation below 0.1\% level.

\section{Analysis and results}
\label{sec:analysis}

\label{dataset}
\begin{figure}
\includegraphics[width=0.5\textwidth]{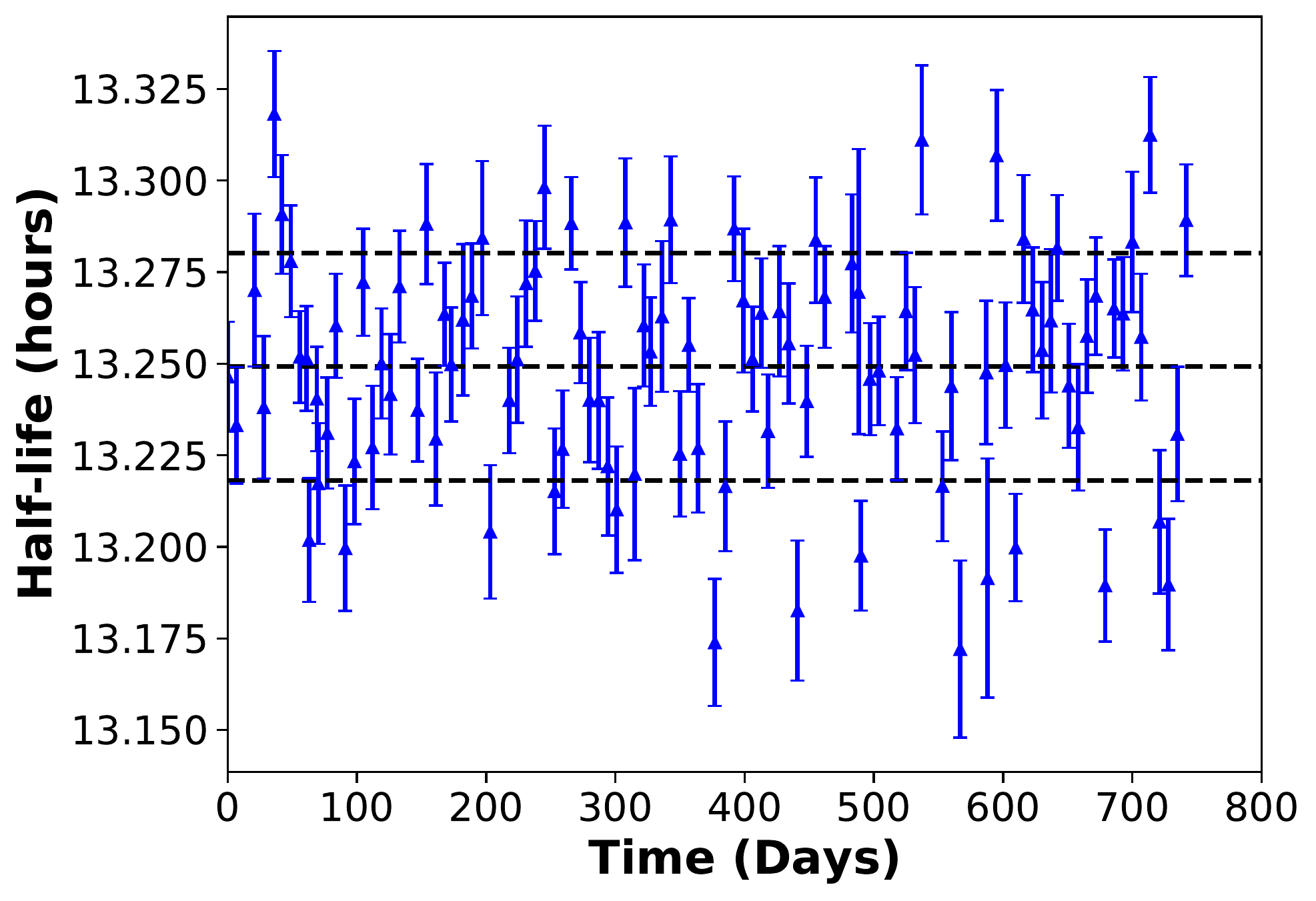}
\caption{Half-life time-series (along with error bars) for \I using the data from B18. The dashed horizontal lines indicate the $\pm 1\sigma$ range for the data.}
\label{fig1}
\end{figure}

\begin{figure}
\includegraphics[width=0.5\textwidth]{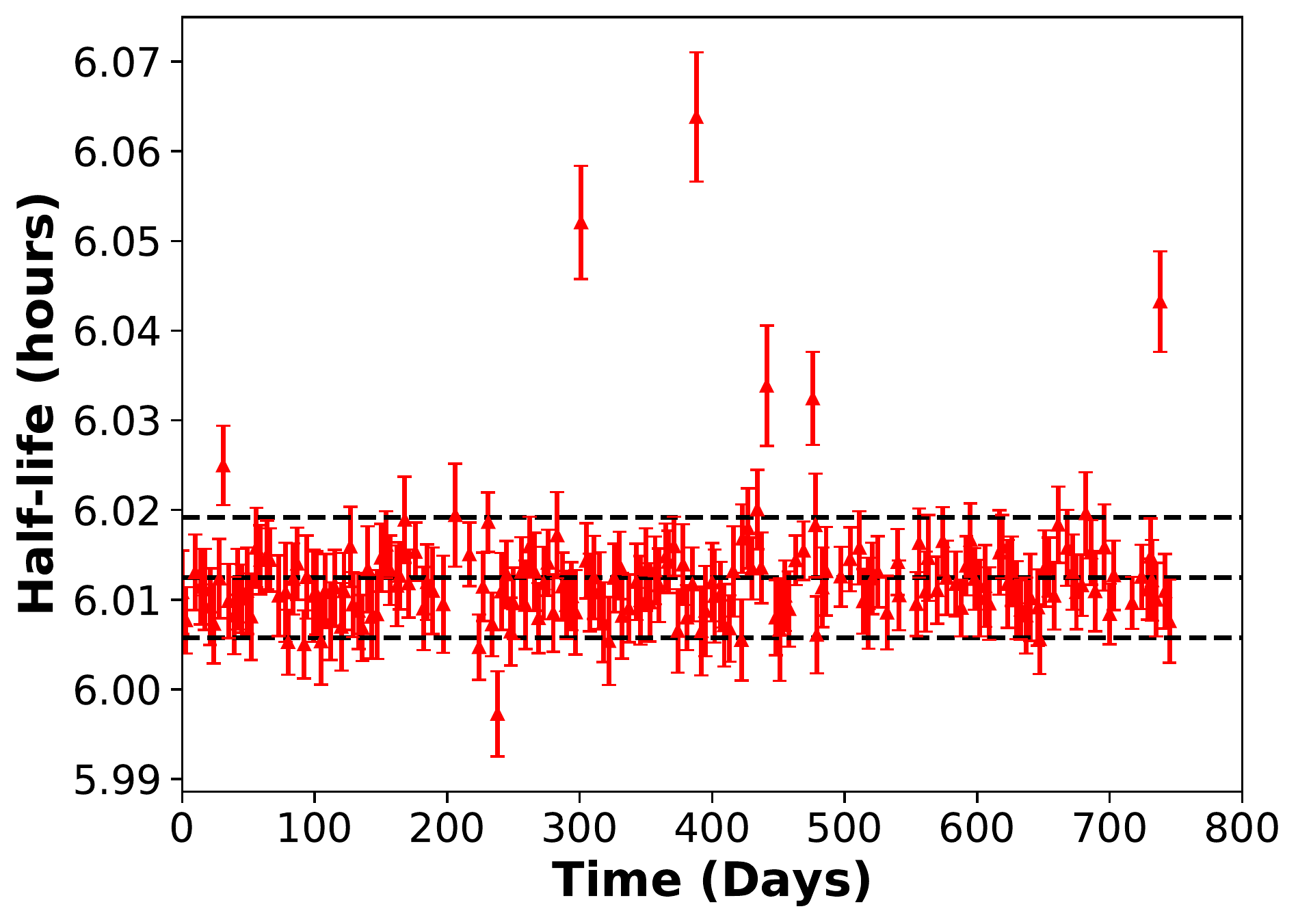}
\caption{Half-life time-series for \tc (along with error bars) using the data from B18. The dashed horizontal lines indicate the $\pm 1\sigma$ range for the data.}
\label{fig2}
\end{figure}

The \I decay data  comprise  101 measurements, of which one was discarded because of experimental disturbances. Similarly, the  \tc decay data comprise of 186 measurements, of which 11 were discarded because of an error in the sample preparation. Both these sets of decay measurements  along with the associated errors were kindly made available to us by Dr. Borrello. The outliers were already removed from the dataset, so no additional pruning had to be done. These measurements are plotted in Fig.~\ref{fig1} and Fig.~\ref{fig2}  for \tc and \I, respectively.

\begin{figure}
\includegraphics[width=0.5\textwidth]{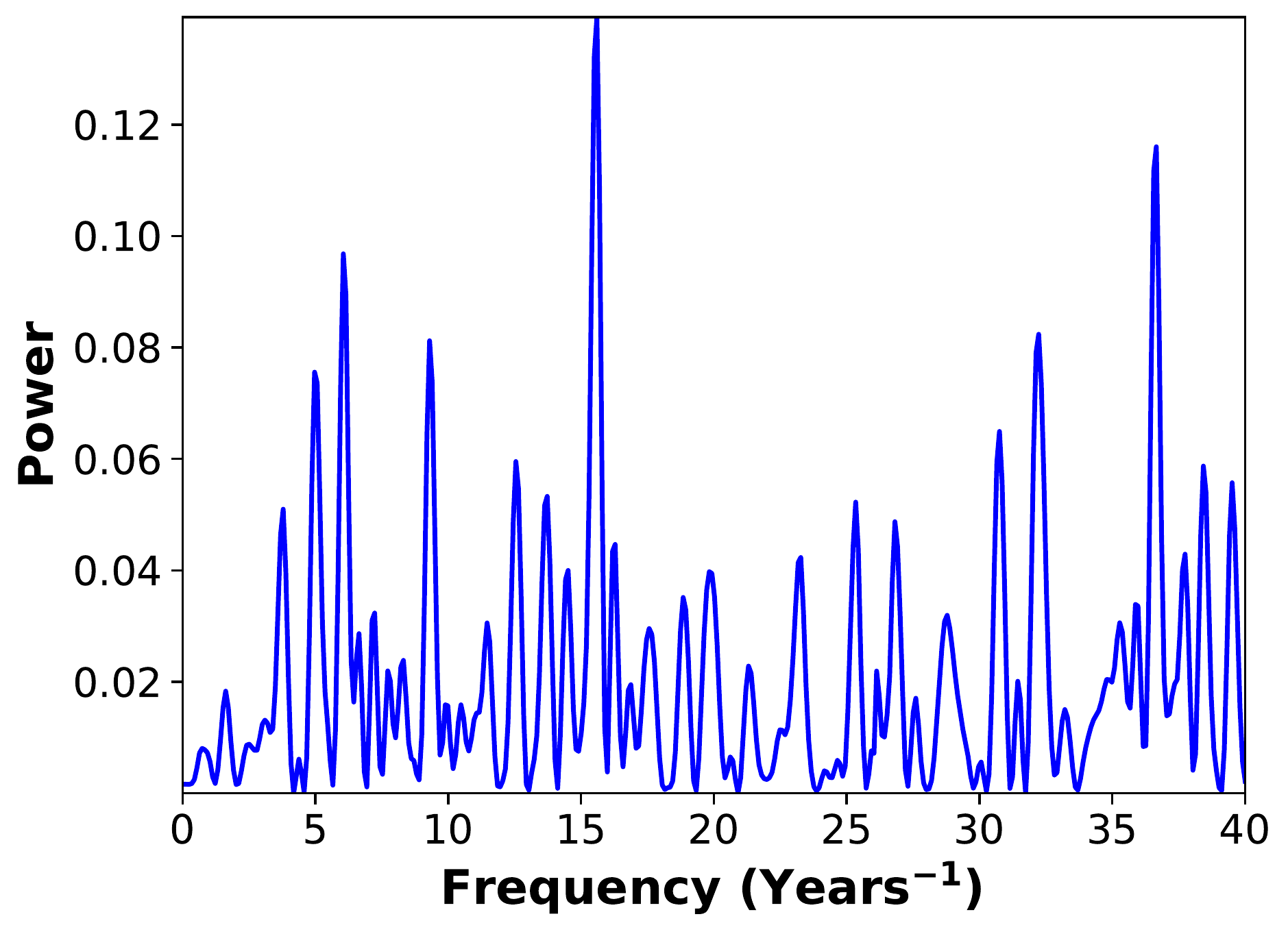}
\caption{Generalized L-S periodogram for \I shown upto  frequency of 40/year. We also searched for statistically significant peaks at higher frequencies, upto 123/year, but did not find any.}
\label{LS:I}
\end{figure}

\begin{figure}
\includegraphics[width=0.5\textwidth]{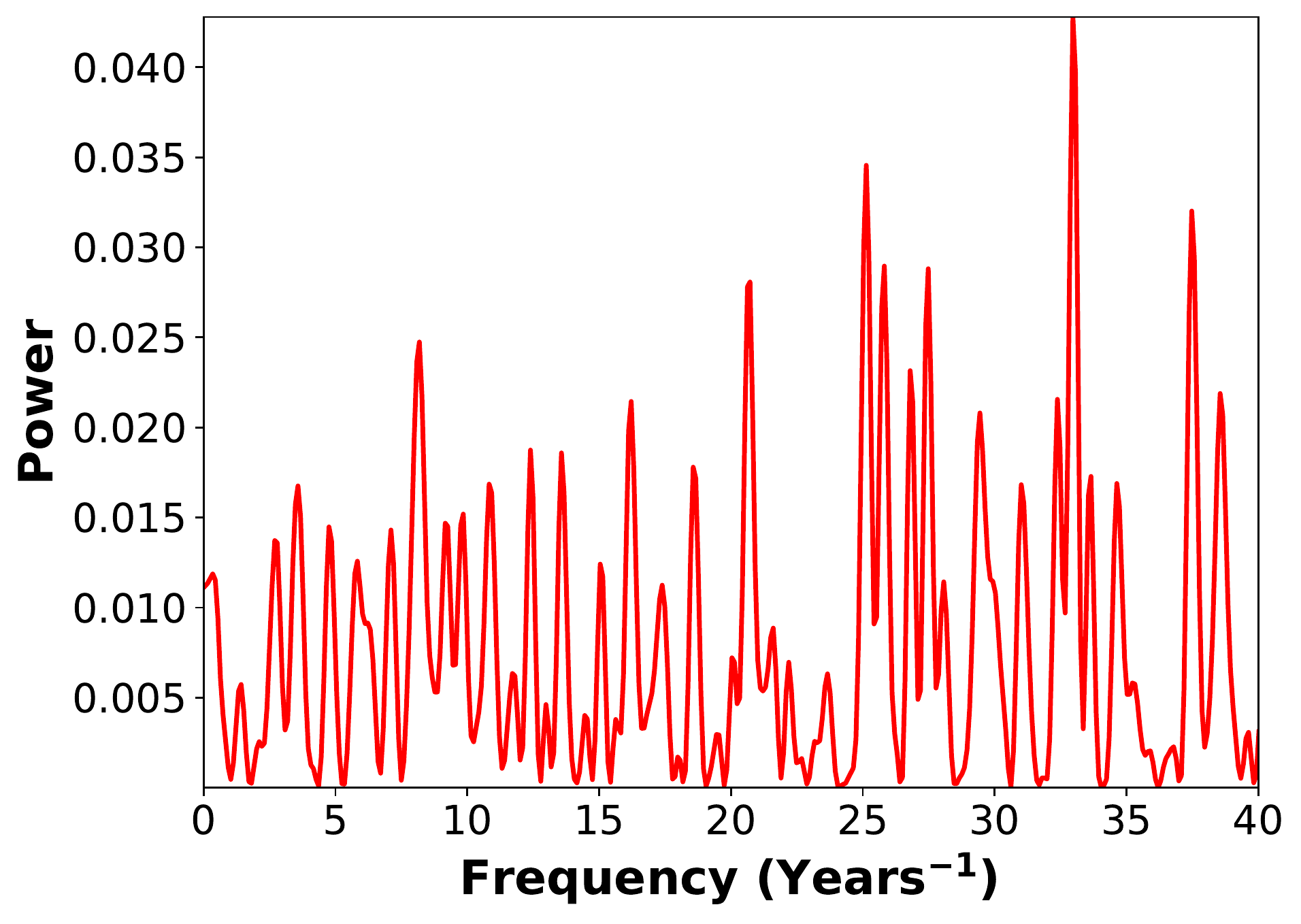}
\caption{Generalized L-S periodogram for \I shown upto  frequency of 40/year. We also searched for statistically significant peaks at higher frequencies, upto 123/year, but did not find any.}
\label{LS:Tc}
\end{figure}

We now apply generalized L-S periodogram to this dataset. We used the L-S implementation in {\tt astropy}~\cite{astropy}. For the frequency resolution and maximum frequency needed for the L-S analysis, we followed the recommendation in ~\citet{Vanderplas}; viz. the size of each frequency bin is the reciprocal of five times the total duration of the dataset, and the maximum frequency is equal to five times the mean Nyquist equivalent frequency. Therefore, for \I the frequency resolution is equal to 0.000269 day$^{-1}$ (0.098 year$^{-1}$) and maximum frequency equal to 0.337 day$^{-1}$ (123 year$^{-1}$). For \tc, the corresponding numbers are   0.000268 day$^{-1}$ (0.098 year$^{-1}$) and 0.620 day$^{-1}$ (226.3 year$^{-1}$), respectively. However, since the astrophysically interesting frequencies are at 1/year and 8-14/year (associated with solar rotation)~\cite{Sturrockcomp,Sturrocksolar}, for brevity we only display the L-S periodogram upto a maximum range of 40/year. However, we also checked that there are no significant peaks at higher  frequencies.   We normalized the periodogram by the residuals of the data around the constant reference model. With this normalization, the L-S power varies between 0 and 1. This is similar to the normalization used in ~\cite{Tejas,Dhaygude}. On the other hand, B18  (also~\cite{Desai16}) used the normalization proposed by Scargle~\cite{Scargle}. The relation between these two normalizations  is outlined in ~\cite{Tejas,Vanderplas}. 
The plots showing the L-S periodograms for \I and \tc can be found in Fig.~\ref{LS:I}
and Fig.~\ref{LS:Tc}, respectively. There are no huge peaks which stand out in these periodograms. Therefore, we find  that our more sensitive method of looking for periodicities using the generalized L-S periodogram also does not reveal any significant peaks.
However, we also quantify this by formally calculating the FAP using all the different methods outlined in Sect.~\ref{sec:ls}. The L-S powers, FAP (using all these methods) for the frequency with the maximum power, frequency associated with solar rotation, as well as that  for annual modulation are shown in Table~\ref{tab:Tc} and Table~\ref{tab:I} for \tc and \I, respectively.  For \tc, the maximum power is seen at a period of about 11 days, with FAP (using bootstrap method) of about 0.9.
For \I, the maximum peak is seen at 23.39 days, with FAP (using the ``Naive'' method) of about 0.14. This corresponds to a $Z-$score of  only 1.1$\sigma$, computed using the prescription in ~\cite{Cowan11,Ganguly}.
As we can see, none of the FAPs are smaller than 0.05, and the FAP for frequencies associated with solar rotation  as well as annual modulation are greater than 0.1.

Therefore, we concur with B18 that there are no periodicities in the nuclear decay rates for \I and \tc using the two year data accumulated at the Bronson Methodist hospital.



\vspace{3mm}
\begin{table*}
\begin{tabular}{|c|c|c|c|c|c|c|}

 \hline
 Period (Days) & Frequency (year$^{-1}$) & L-S Power & FAP : Baluev  & FAP : Davies  &FAP : Naive    & FAP : Bootstrap  \\
 \hline

11.1 & 32.96 & 0.043 & 1.0 & $>1.0$ & 0.99 &	0.920\\


354.8 & 1.03 & 0.00046 & 1.0 & $>1.0$ & 1.0 &	1.0\\
 
44.6 & 8.18 & 0.025 & 1.0 & $>1.0$ & 1.0 &	1.0\\


 \hline
\end{tabular}
\caption{ \textbf{\tc}  L-S powers and FAP for our data using multiple methods:  Baluev,  Davies, Naive, and Bootstrap. We show the corresponding values of the period and frequency for the most significant peak (corresponding to the period of 11.07 days), followed by the period closest to the annual variation (365 days), as  well as the period with maximum power in the  solar rotation range (44.61 days).  As we can see, all the FAPs are close to 1, and hence not significant.}
\label{tab:Tc}
\end{table*}

\vspace{3mm}
\begin{table*}
\begin{tabular}{|c|c|c|c|c|c|c|}

 \hline
 Period (Days) & Frequency (year$^{-1}$) & L-S Power & FAP : Baluev  & FAP : Davies  &FAP : Naive    & FAP : Bootstrap  \\
 \hline

23.4 & 15.59 & 0.14 & 0.38 & 0.48 & 0.14 &	0.26\\


353.2 & 1.03 & 0.0056 & 1.0 & $>1.0$ & 1.0 &	1.0\\
 
39.2 & 9.3 & 0.081 & 0.99 & $>1.0$ & 0.98 &	0.998\\


 \hline
\end{tabular}
\caption{ \textbf{\I}  L-S powers and FAP for our data using multiple methods:  Baluev,  Davies, Naive, and Bootstrap. Similar to Table~\ref{tab:Tc}, we find the corresponding values for the period and frequency of the most significant peak  (corresponding to the period of 23.39 days), followed by the period closest to the annual variation (365 days), as  well as the period with maximum power in the  solar rotation range (39.24 days).  As we can see, all the FAPs are $>$0.1, and are hence not significant.  The peak with the maximum power (at 23.39 days) has the FAP of 0.14, corresponding to the $Z$-score of only 1.1$\sigma$.}
\label{tab:I}
\end{table*}


\section{Conclusions}
\label{sec:conclusions}
The aim of this work was  to carry out an independent analysis  of the \tc and \I nuclear decay rates, to look for statistically significant periodicities at frequencies, for which cyclic modulations have previously been found using other nuclei. The nuclear decay measurements were carried out in the Nuclear Medicine department at the Bronston Methodist Hospital in Michigan and are described in further detail in B18. \tc and \I decay by isomeric transitions and electron capture, respectively. \rthis{Prior to this work, there were no searches for periodicities from nuclei with isomeric transitions, and   only one search in case of electron capture.} 

For this purpose,  we used the  generalized or floating-mean  L-S  periodogram~\cite{Kurster} (similar to our previous works~\cite{Desai16,Tejas,Dhaygude}), as it is more sensitive than the ordinary \mbox{L-S} periodogram, which was used in B18. We searched for statistically significant peaks for both these nuclei upto five times the Nyquist frequency. This frequency range encompasses the band from 8 to 14  per year (which could contain signatures of influence from solar rotation) and also the annual modulation (in case of any influence due to the Earth-Sun distance). 

The generalized L-S periodograms (upto a frequency range of 40/year) can be found in Fig.~\ref{LS:I} and Fig.~\ref{LS:Tc}. The FAP for the highest peak,  the frequency closest to one year, and also for the frequency with highest FAP between 8-14 per year can be found in Table~\ref{tab:Tc} and Table~\ref{tab:I}. We do not find statistically significant peaks at any of these frequencies and the FAP for the peak with highest power is close to 1, indicating there is no  periodicity at any frequency.

To promote transparency in data analysis, we have made our analysis codes and data available online, which  can be found at \url{https://github.com/Gautham-G/Lomb-Scargle-Analysis}. 

\begin{acknowledgements}
We are grateful to Joseph Borrello for providing us the data from B18 used for this analysis.
\end{acknowledgements}

\bibliography{lombscargle}

\end{document}